\documentclass[twocolumn,preprintnumbers,amsmath,amssymb,superscriptaddress]{revtex4}
\usepackage{graphicx}
\usepackage{dcolumn}
\usepackage{bm}
\usepackage{natbib}

\def\ltsima{$\; \buildrel < \over \sim \;$}
\def\simlt{\lower.5ex\hbox{\ltsima}}
\def\gtsima{$\; \buildrel > \over \sim \;$}
\def\simgt{\lower.5ex\hbox{\gtsima}}

\begin{document}
\title{An Alternative Approach to Holographic Dark Energy}

\author{Fergus Simpson}
\email{frgs@roe.ac.uk} \affiliation{Institute of Astronomy,
University of Cambridge, Madingley Road, Cambridge CB3 0HA}
\affiliation{Institute for Astronomy, University of Edinburgh,
Royal Observatory, Blackford Hill, Edinburgh EH9 3HJ}

\date{\today}
\newcommand{\ud}{\mathrm{d}}

\begin{abstract}
Here we consider a scenario in which dark energy is associated
with the apparent area of a surface in the early universe. In
order to resemble the cosmological constant at late times, this
hypothetical reference scale should maintain an approximately
constant physical size during an asymptotically de-Sitter
expansion. This is found to arise when the particle horizon -
anticipated to be significantly greater than the Hubble length -
is approaching the antipode of a closed universe. Depending on the
constant of proportionality, either the ensuing inflationary
period prevents the particle horizon from vanishing, or it may
lead to a sequence of ``Big Rips".
\end{abstract}
\maketitle

\section{Introduction}

The acceleration of our universe is now a well established
phenomenon, and numerous theoretical approaches have attempted to
provide an explanation. The cosmological constant is arguably the
strongest candidate, although it has a number of well-documented
issues, particularly with regard to its size. One of the more
attractive methods of attaining such a small number is by relating
it to a cosmological length scale.

The  holographic principle speculates that all information stored
within some  volume  is  represented on the boundary of that
region.  A cosmological variation of the holographic principle was
originally proposed by Fischler \& Susskind \cite{fischler-1998-}.
In this scenario the cosmological horizon acts as a surface which
obeys the Bekenstein bound, limiting the entropy of the enclosed
volume, such that
\begin{equation} \label{eq:bek}
S \leq \frac{A}{4G} .
\end{equation}
This concept has been extended by Bousso \cite{bousso-1999-9907}
into a covariant and more general conjecture.

Cohen et al. \cite{cohen-1999-82} took a slightly different
approach, whereby the dark energy density is proportional to the
inverse square of some cosmological length. This arises when
constraining the energy in some volume to be less than a black
hole of the same size.

\begin{equation} \label{eq:cohen}
L^3 \rho_\Lambda \lesssim L M_P^2
\end{equation}

\noindent The most natural choices for $L$ are the particle
horizon, and the Hubble length. However Hsu \cite{hsu-2004-594}
has shown that under adiabatic conditions, neither exhibit the
equation of state necessary to mimic dark energy. One could
enforce $w=-1$, by permitting energy exchange between the dark
matter and dark energy, however this approach has not proved
particularly successful \cite{wang-2005-22}.

Li \cite{li-2004-603} produced a model of holographic dark energy
based on Cohen's approach, which has subsequently been explored in
further detail
\cite{setare-2006-0609,2005CQGra..22.4895G,2004JCAP...08..013H,2006GReGr..38.1285N}
and is compatible with current observational constraints
\cite{zhang-2005-72}. By identifying $L$ as the future event
horizon, the required equation of state could emerge without
invoking energy exchange. For a saturated inequality, the energy
density is given by

\begin{equation} \label{eq:holo}
\rho_\Lambda = 3 d^2 M_p^2 L^{-2} .
\end{equation}

\noindent The value $d=1$ is usually adopted for the reasons
outlined in \cite{li-2004-603} - in particular, $d \geq 1$ ensures
the entropy doesn't decrease. However, within Li's model there
remain two areas of concern. First of all, for self-consistency,
there is the requirement that all forms of energy must decay at
some time in the future. And secondly, the initial condition of a
preset future event horizon appears a little unusual, raising the
issue of causality.

Here we take a rather different approach, finding that a more
natural form of dark energy may arise when considering the
particle horizon. We argue that its associated area may possess
greater physical significance than the length scale itself. In
Section \ref{sec:phor} we explore the consequences of attributing
an energy density to the area of the particle horizon within a
closed universe. Section \ref{sec:eqstat} establishes the equation
of state, and explores different values of the constant of
proportionality $d$. In some cases this may lead to a ``Big Rip"
scenario. Section \ref{sec:obcons} summarises the current and
future observational constraints on the model. We speculate on the
possible physical interpretations in Section \ref{sec:physmot}.

\section{Particle Horizon} \label{sec:phor}

Here we consider the consequences of ascribing an energy density
to the apparent area of the particle horizon (which represents the
intersection of our past light cone with a particular point in
time prior to inflation). Our parameterisation of the
proportionality constant, $12 \pi d^2$, retains the close
relationship with (\ref{eq:holo}) and will simplify our analysis
in the following section.

\begin{equation} \label{eq:area}
\rho_\Lambda = \frac{12 \pi d^2 M_p^2}{A}
\end{equation}

\begin{figure}
\begin{center}
\includegraphics[width=3in]{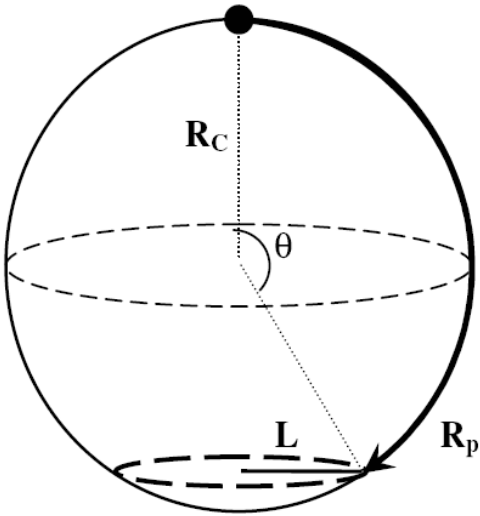}
\end{center}
\caption[The spread of the particle horizon across a closed
universe]{\label{fig:holo}The spread of the particle horizon $R_P$
across a closed universe, with the radius of curvature $R_C$ and
corresponding angle $\theta$ also labelled. Whilst the comoving
area shrinks once beyond the half-way point, the physical area of
the particle horizon never decreases provided $d \geq 1$. The
residual distance is effectively the future event horizon.}
\end{figure}

\noindent The area $A$ associated with our past light cone in a
closed universe is given by

\begin{equation} \label{eq:area2}
A = 4 \pi R_C^2 \sin^2 \theta ,
\end{equation}

\noindent where $\theta = R_P/R_C$, as illustrated in Fig.
\ref{fig:holo}, and the particle horizon $R_P$ is

\begin{equation}\label{eq:phorizon}
R_P = a(t) \int_0^{t} \frac{\ud t'}{a(t')} ,
\end{equation}

\noindent and $R_C = R_0 a = H^{-1}\Omega_k^{-1/2}$ is the radius
of curvature.

The vanishing particle horizon within a closed universe has caused
some concern \cite{fischler-1998-,kaloper-1999-60} but since we
are attributing its (inverse) size to the dark energy density, we
find that this may produce cosmic acceleration, so the recession
is strongly suppressed, and the entropy bound is protected
(provided $d \geq 1$).

\section{The Equation of State} \label{sec:eqstat}

Taking the time derivative of (\ref{eq:area}), and utilising the
relations

\begin{equation}
\dot{\rho} = -3\frac{\dot{a}}{a} \rho (1+w)
\end{equation}

\noindent and

\begin{equation}
\dot{R}_P = 1  + H R_P
\end{equation}

\noindent which arose from differentiating (\ref{eq:phorizon}), we
find that the equation of state can be expressed as a function of
$\Omega_\Lambda = \rho_\Lambda/\rho_c$.

\begin{equation} \label{eq:fullw}
w = -\frac{1}{3} + \frac{2}{3} \frac{\sqrt{\Omega_\Lambda}}{d}
\cos{\theta} .
\end{equation}

This clarifies the behavior of $w$ as the particle horizon
traverses the closed universe. In the following subsections we
explore the significance of $d$, assuming that it exhibits no
significant time-variation.

\subsection{The case $d=1$: De Sitter}

For the scenario in which $d=1$ there are three distinct regimes

\begin{eqnarray}
\label{eq:array1}w = +\frac{1}{3} & (\Omega_\Lambda = 1, R_P \ll R_C) \\
\label{eq:array2}w = - \frac{1}{3} & (\Omega_\Lambda = 0) \\
\label{eq:array3} w = - 1  &  (\Omega_\Lambda = 1, R_P \simeq \pi
R_C)
\end{eqnarray}

Both (\ref{eq:array1}) and (\ref{eq:array2}) will ultimately
converge to (\ref{eq:array3}). For this limiting case, we have
$|\Omega_k| \ll 1$, and the particle horizon is in close proximity
to the observers antipode. Then our expression for the energy
density (\ref{eq:area}) simplifies to

\begin{equation} \label{eq:areaflat}
\rho_\Lambda = \frac{d^2}{(\pi R_C - R_P)^2}
\end{equation}

\noindent for $ R_P/R_C \simeq \pi $. This leads to

\begin{equation} \label{eq:w}
w = - \frac{1}{3} - \frac{2}{3} \sqrt{\Omega_\Lambda} .
\end{equation}

Note that having set $d=1$, the evolution of the equation of state
matches that of Li's approach, since the particle horizon is
identified with the future event horizon (see Fig.
\ref{fig:holo}). While the future event horizon was a free
parameter in Li's model, here it is dictated by global geometry,
which in turn was determined by the suppression of $\Omega_k$
during inflation. Once inflation has been established, the
particle horizon is frozen in at some fraction of the radius of
curvature. Further growth of $\theta$ occurs during radiation and
matter domination, scaling as $a$ and $\sqrt{a}$ respectively.

Given an observational fit of the form $w=w_0+w_a(1-a)$, and
taking $\Omega_\Lambda = 0.75$, one expects

\begin{equation} \label{eq:wz}
w \simeq -0.9 + 0.2 (1-a) .
\end{equation}

\subsection{The case $d>1$: Decay}

For greater values of $d$, there is initially a higher energy
density, and thus acceleration arises earlier. This ensures
$\theta < \pi$ is sustained, although we still find $\theta$
converging to $\pi$ for all values of $d$. The only significant
difference from the $d=1$ case is the gradual decay of the vacuum
energy. Again, for small $\Omega_k$ we have

\begin{equation} \label{eq:wdecay}
w = - \frac{1}{3} - \frac{2}{3} \frac{\sqrt{\Omega_\Lambda}}{d} .
\end{equation}

\subsection{The case $d<1$: Big Rips}

We can see from (\ref{eq:fullw}) that $d<1$ may result in a
phantom cosmology, characterised by $w<-1$. This leads to a
divergence in the energy density, destroying all galactic and
atomic structure within a finite time \cite{2003PhRvL..91g1301C}.
In this context, the value of the Hubble parameter is initially
too low, so the particle horizon is not prevented from vanishing.
Let us take $d=0.5$ as an illustration. This provides an equation
of state asymptotically approaching $w=-5/3$, and so $\rho_\Lambda
\propto a^2$.  The unusual scenario here is that at some point -
which we take to be the Planck density - we pass $\theta = \pi$,
and there is a radical transition in behaviour.  The area of the
particle horizon begins to increase in size once more, returning
to its point of origin, and we find $w=1$. We can then establish a
series of ``Big Rips", each taking much longer than its
predecessor, and with the number of e-folds dictated by $d$. For
instance, with $d=0.5$ a present-day value of $\rho_\Lambda \sim
10^{-120}$ implies a further $\sim 140$ e-folds until the Planck
density is reached. If $d=0.8$, then $\rho_\Lambda \propto
\sqrt{a}$, so a further $\sim 550$ e-folds are required.  Note
that $d<0.5$ appears to be forbidden on the grounds that this
could lead to an unphysical value of $w>1$.

\subsection{Entropy}

Could the particle horizon still obey the constraint given by
(\ref{eq:bek})? As explored by Fishler \& Susskind, the most
natural way to count the entropy enclosed within a given area is
that which has passed though the past light cone. The most
important consequence of this is that we bypass the vast volume of
the particle horizon associated with inflation, as we need only
consider the light cone after reheating.  Thus one could
potentially reconcile a large universe with the constraint given
by (\ref{eq:bek}). This bound is eventually violated when
considering the $d<1$ models. However, the significance of this is
questionable, since a simple model of a collapsing universe would
also lead to a violation.

\subsection{Theoretical Issues}

One of the problems associated with dark energy is the question of
coincidence. Why are we around at the time of transition from
matter to dark energy domination? There are few scenarios which
can genuinely address this problem. One is where recent events -
such as the formation of structure - are directly related to
cosmological kinetics. Efforts to explain dark energy via the
gravitational influence of nonlinearities have thus far proved
unsuccessful. The second is one in which the lifetime of the
universe is restricted. If either recollapse or the big rip occurs
within the next $\sim 10^{12}$ years then we are no longer
unusually close to the time of transition. This makes studying
cosmologies with $w<-1$ particularly attractive, despite the
difficulty in interpreting the thermodynamics.

The standard cosmological constant also runs into difficulty when
we try to establish why the energy density appears so small.
Conversely, here we must wonder why the energy density is not
\emph{even smaller}. If there were a few extra e-folds of
inflation this would generate a much larger volume within which we
would be more likely to exist. This could be alleviated if other
important factors, such as the magnitude of primordial density
perturbations, were also correlated with the number of e-folds of
inflation.

\section{Observational Constraints} \label{sec:obcons}

Since the equation of state with small $\Omega_k$, given by
(\ref{eq:wdecay}),  matches earlier studies using the future event
horizon, we can utilise constraints on $d$ from Chang et al.
\cite{chang-2006-633}. They find an equation of state of the form
(\ref{eq:wdecay}) remains compatible with current observational
data from supernovae, the CMB, and large scale structure. There is
however a preference for $d<1$, a value which as we have seen
leads to some intriguing behaviour. They also highlight how
constraints are particularly sensitive to the Hubble parameter,
with a lower $h$ favouring higher $d$.

Recently, it has been suggested that the controversial
Heidelberg-Moscow claim of neutrinoless double beta decay
\cite{kleingrothaus-2004-586} provides an indication of a
phantom-like dark energy, with $w<-1$
\cite{macorra-2006-,2005PhRvL..95v1301H}. However, it should be
noted that even if the neutrino result is verified,  the constant
$w$ parameterisation can provide misleading results
\cite{2002PhRvD..65l3003M,2006PhRvD..73h3001S}, with an evolving
equation of state mimicking the phantom behaviour. This leaves us
with the unusual situation whereby evidence for $w<-1$ could still
suggest $w(z)>-1$.

\section{Physical Motivation} \label{sec:physmot}

A potential relationship between cosmological length scales and
dark energy has received a great deal of attention, with limited
physical justification. Here we assess the possible
interpretations and motivations for associating the particle
horizon with the vacuum energy. At first glance, the area of the
particle horizon simply corresponds to the surface of infinite
redshift surrounding us. However this may take on greater physical
significance if there was a critical change in the underlying
physics in the very early universe. This would correspond to a
particular point in time prior to inflation. We would then sample
this `horizon' surface which intersects our past light cone, in
much the same way as our view of the last scattering surface
samples a preferential scale. Properties of the present day
universe may then be influenced by this reference length scale.
This could be related to the proposal by Padmanabhan
\cite{padmanabhan-2005-22} in which vacuum fluctuations within
finite regions of space are responsible for the cosmological
constant.

This work was also partly motivated by the following prospect. If
the universe is found to be slightly closed, with $\Omega_k \sim
0.01$, we would be pushed deeper into the coincidence problem.
There appears little reason to believe the curvature should be
comparable to the matter and dark energy density. However if there
was some relationship between the global curvature and the dark
energy, this would partly alleviate the problem. Unfortunately we
cannot make quantitative predictions on $\Omega_k$ here, since it
present value will depend on the nature of inflation.

\section{Discussion} \label{sec:discholo}

An energy density associated with the size of the particle horizon
within a closed universe has been shown to demonstrate a negative
equation of state. Precise behaviour depends on the constant of
proportionality, $d$. Whilst $d=1$ is the most natural choice,
imitating the cosmological constant at late times, current
observational constraints slightly prefer $d<1$. This corresponds
to a cosmology with a sequence of ``rips", as the particle horizon
repeatedly traverses the closed universe. Such a model has the
advantage of addressing the coincidence problem, and has a simple
mechanism for lowering the equation of state at late times. We
have also established a robust lower bound of $d>0.5$, which must
be satisfied in order to ensure $w<1$ in the early universe.

For the case $d=1$ we find cosmic evolution equivalent to the
model put forward by Li, which was based on associating the dark
energy density with the future event horizon. However by using the
particle horizon, this approach is causal, does not require the
decay of matter, and has a natural reference scale - the radius of
curvature.

As with other holographic models, this approach is rather
speculative, but does feature a number of attractive properties.
Given the close relationship predicted between $\Omega_\Lambda$
and $w$, data from future dark energy surveys such as SNAP should
be capable of distinguishing this model from the cosmological
constant.

\bibliography{dis}
\end{document}